\documentclass{cup-hpl}

\usepackage[utf8]{inputenc}
\usepackage[T1]{fontenc}
\usepackage[nolist]{acronym}
\usepackage{textgreek,ulem,xcolor}
\usepackage{amsmath}
\usepackage{graphicx}
\usepackage{verbatim}
\usepackage{bm}
\usepackage[nolist]{acronym}
\usepackage{hyperref}
\usepackage[numbers]{natbib}
\usepackage{ulem}
\usepackage{color, soul}
\soulregister\ac7
\usepackage{multicol}
\usepackage{float}

\DeclareUnicodeCharacter{2009}{\,}

\newcommand{\correction}[2]{#2}                    

\begin{document}

\shorttitle{Apollon Real-Time AO}                                   
\shortauthor{J.B. Ohland et al.}
	
\title{Apollon Real-Time Adaptive Optics (ARTAO) - Astronomy-Inspired Wavefront Stabilization in Ultraintense Lasers}

\author[1,2]{J. B. Ohland \corresp{J.B. Ohland \email{j.b.ohland@gsi.de}}}
\author[1]{N. Lebas}
\author[3]{V. Deo}
\author[3,4,5,6]{O. Guyon}
\author[1]{F. Mathieu}
\author[1]{P. Audebert}
\author[1]{D. Papadopoulos}

\address[1]{Laboratoire d’Utilisation des Lasers Intenses, Ecole Polytechnique, 91128 Palaiseau Cedex, France}
\address[2]{GSI Helmholtzzentrum für Schwerionenforschung, Planckstraße 1, 64291 Darmstadt, Germany}
\address[3]{Subaru Telescope, National Astronomical Observatory of Japan, National Institute of Natural Sciences, Hilo, HI, U.S.A.}
\address[4]{Astrobiology Center of NINS, 2-21-1, Osawa, Mitaka, Tokyo, Japan}
\address[5]{Steward Observatory, University of Arizona, Tucson, AZ, U.S.A.}
\address[6]{College of Optical Sciences, University of Arizona, Tucson, AZ, U.S.A.}

\begin{abstract}
    Traditional wavefront control in high-energy, high-intensity laser systems usually lacks real-time capability, failing to address dynamic aberrations. This limits experimental accuracy due to shot-to-shot fluctuations and necessitates long cool-down phases to mitigate thermal effects, particularly as higher repetition rates become essential, e.g. in Inertial Fusion research.\\
    This paper details the development and implementation of a real-time capable adaptive optics system at the Apollon laser facility. Inspired by astronomical adaptive optics, the system uses a fiber-coupled 905\,nm laser diode as a pilot beam that allows for spectral separation, bypassing the constraints of pulsed lasers. A GPU-based controller, built on the open-source CACAO framework, manages a loop comprising a bimorph deformable mirror and high-speed Shack-Hartmann sensor. Initial tests showed excellent stability and effective aberration correction. However, integration into the Apollon laser revealed critical challenges unique to the laser environment that must be resolved to ensure safe operation with amplified shots.
\end{abstract}

\maketitle

\section{Introduction}
    Semi-static \ac{AO} has been an essential tool in controlling beam quality in high-energy, high-power laser facilities. These systems effectively remove static aberrations, enabling the highest intensities on target and unlocking new experimental regimes\cite{druon1998, baumhacker2002, wattellier2003, samarkin2016, ohland2022}. With the rapid advancement of laser technology, many beam quality issues have been addressed, revealing limitations that were previously unnoticed or considered irrelevant. Among these newly recognized limitations, dynamic aberrations have gained attention in recent years \cite{ohland2023, negro2018, leroux2018, ohland2022diss}.\\
    Dynamic aberrations evolve over various timescales: they range from slow drifts, e.g. due to temperature changes of optical tables, over medium timescale changes, such as cooldown effects of active media, up to fast evolution due to air turbulence or system vibrations. While pointing jitter is often mitigated using fast tip-tilt mirrors, higher-order aberrations have not yet been successfully addressed in real time. The increasing public interest in high-energy lasers with higher repetition rates, as demonstrated by initiatives like the EU-funded \ac{THRILL} program\footnote{https://www.thrill-project.eu/} and developments towards \ac{IFE}, underscores the urgency of addressing this issue.\\
    Given the limitations of passive mitigation strategies for dynamic aberrations, the path forward clearly involves active beam control through \ac{RTAO}\footnote{Please note that the term "\ac{AO}" usually includes the real-time definition. However, in the laser community, "\ac{AO}" referred to non-real-time systems for the past decades, which is why we deliberately chose use the "\ac{RTAO}" abbreviation to distinguish these systems in our context.}. This technology has been successfully applied for real-time wavefront corrections in fields such as astronomical instrumentation\cite{jovanovic2015}, ophthalmology\cite{akyol2021}, and free-space optical communication\cite{guiomar2022, kudryashov2018}. However, the high-energy laser community faces a significant challenge: due to the historical focus on (semi-)static \ac{AO}, there is a lack of expertise in designing and constructing custom \ac{RTAO} systems tailored to the specific needs of large high-power laser facilities. Consequently, no feasible commercial solutions currently exist.\\
    In this work, we take the first step towards addressing this issue by implementing a potential \ac{RTAO} solution for a specific case. This implementation provides valuable insights into system architecture, feasible technologies, and field-specific challenges. These insights will lay the foundation for the development of readily available \ac{RTAO} systems in the field of scientific high-energy, high-power lasers, paving the way for a new era of beam stability in our community.\\
    \newline
    This paper is structured as follows: we begin with an introduction to the situation at the Apollon Laser in France in the next section, which was the motivation and use case that we focused on. We proceed with section \ref{sec:design-considerations} with a detailed description of the design considerations for our \ac{RTAO} system, as well as the specific implementation in section \ref{sec:artao}. We then present the performance evaluation on a test bench in section \ref{sec:characterization} and in the actual system in section \ref{sec:integration}, while we discuss the main challenges that we encountered in the operational environment in detail in section \ref{sec:future-work}. Based on these findings, we outline the future work necessary to integrate \ac{RTAO} into regular operation in high-energy laser systems.

\section{Background} \label{sec:apollon}
    This work was conducted at the Apollon Laser System\cite{burdonov2021, papadopoulos2024}, located close to Paris, France. Apollon is a Titanium Sapphire (Ti:Sapph) \ac{CPA} laser system currently in its ramp-up phase, with the ambitious goal of delivering 150\,J in 15\,fs to the target, achieving 10\,PW peak power at a repetition rate of one shot per minute. The laser pulses are initially generated in an \ac{OPCPA} frontend and then amplified through a series of five bowtie Ti:Sapph multipass amplifiers, each with increasing beam diameter.\\
    \begin{figure}[h]
        \centering
        \includegraphics[width=1\linewidth]{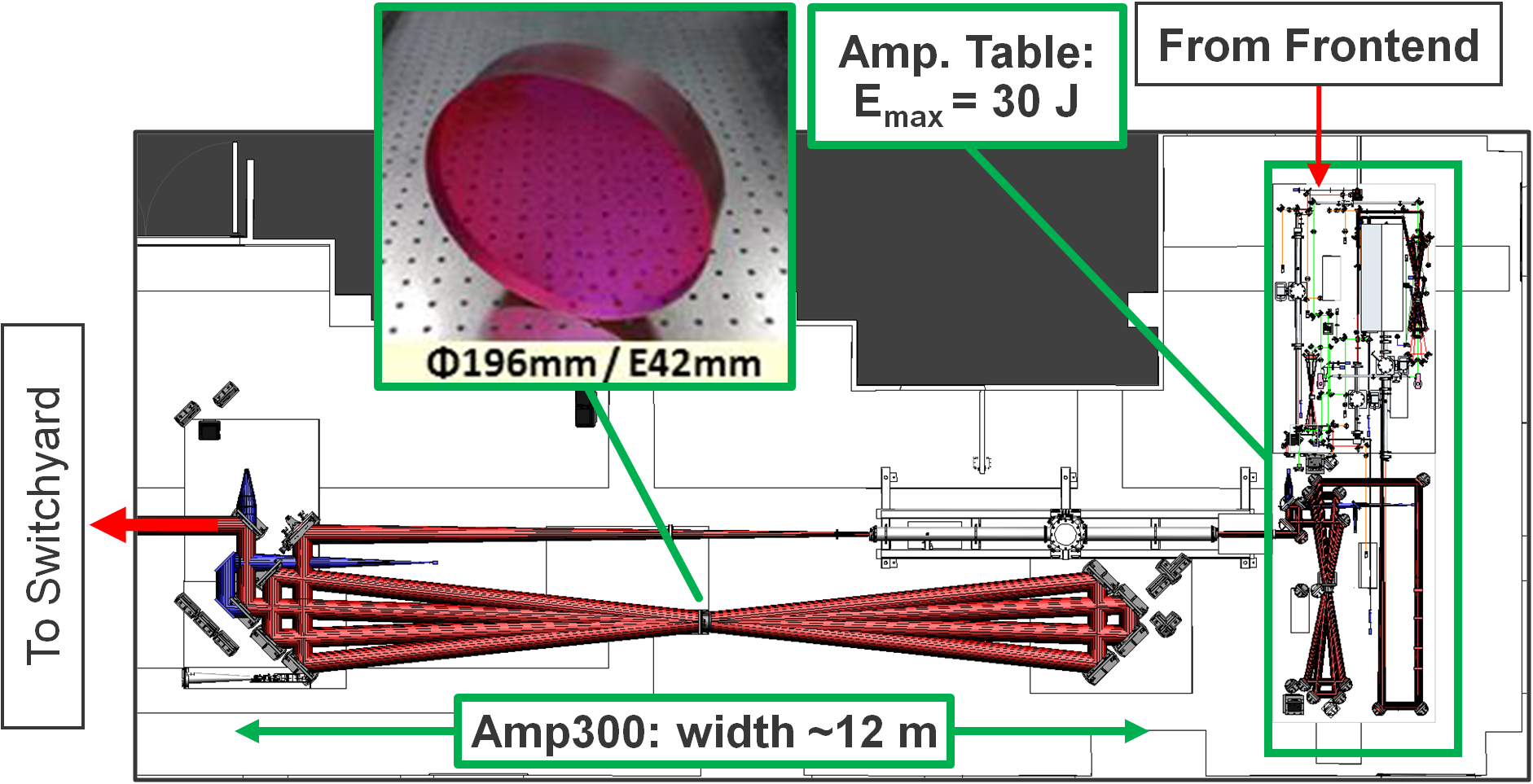}
        \caption{Sketch of the Laser AMplification area (LAM) at Apollon. The last amplifier ("Amp300") was known for causing beam instabilities due to air movement in the beampath.}
        \vspace{-0.3cm}
        \label{fig:apollon-lam}
    \end{figure}
    The final amplifier in this sequence, known as "Amp300," features one of the world's largest Ti:Sapph crystals, measuring 196\,mm in diameter. Amp300 is designed to bring the pulse energy up to 300\,J, with the next upgrade in the pump area within 2024 already unlocking energies of 250\,J. Achieving such high energies is made possible by a large beam diameter of 140\,mm and a complex multiplexing sequence where currently seven pump beams pump the crystal between the four amplification passes.\\
    While this architecture enables large pulse energies, the bowtie design of Amp300, coupled with its large beam diameter, results in over 50\,m of unimaged beam propagation through laboratory air (see Fig. \ref{fig:apollon-lam}). This makes the beam extremely sensitive to air turbulence. Before any countermeasures were implemented, the Strehl ratio fluctuated between 0.2 and 0.9 within a single second, and the local wavefront slope fluctuated at up to 70\,Hz. These fluctuations not only failed to meet the specifications of Apollon (maximum 10\% Strehl ratio fluctuations) but also rendered the use of the Amp300 beamline unfeasible for high-intensity experiments.\\
    Even after implementing passive measures, such as enclosing Amp300 in a tight beam housing, the Strehl ratio fluctuations still exceeded 20\%, highlighting the limitations of passive beam control in certain configurations. This situation underscores the necessity for an active beam control solution, specifically a \ac{RTAO} system.\\
    \newline
    It is important to note that while the architecture of Amp300 makes Apollon particularly susceptible to air turbulence, other facilities face similar challenges that limit certain types of experiments. For instance, the novel ELI NP laser system achieves remarkable beam stability due to its vibration damping system, which includes nearly 1000 springs and dampers supporting the massive 1.5-meter-thick reinforced concrete slab that forms the base of the laser hall. Nevertheless, atmospheric turbulence can still disrupt beam quality sufficiently to prevent certain types of experiments, e.g. when employing helical beams \cite{ohland2023}, which are highly sensitive to angular aberrations \cite{ohland2019}. This issue becomes much more prominent when sudden weather changes such as thunderstorms occur, which cannot be fully compensated by the air-conditioning system.\\
    Furthermore, glass-based high-energy lasers with increased repetition rates are currently being developed for \ac{IFE} applications. The thermal load from these lasers will inevitably generate beam instabilities, further emphasizing the need for expertise in designing, implementing, and operating RTAO systems in. This motivation therefore extends beyond Apollon and permeates the entire high-energy laser community.
    
\section{System Design and Methodology} \label{sec:design-considerations}
    \subsection{Inspiration from Astronomy}
        \ac{RTAO} has been an essential tool in astronomy for decades and is considered obligatory for large, earthbound telescopes in order to overcome the limitations imposed by atmospheric turbulence. Here, various \ac{RTAO} schemes, such as high-resolution Single Conjugate \ac{RTAO} for small fields of view, e.g. for exoplanet imaging\cite{lozi2024}, and Multi Conjugate \ac{RTAO} with multiple \acp{DM} for larger field of view observations\cite{rigaut2018}, have been developed. These systems can handle many hundreds of control modes while operating at crossover frequencies (max. frequency that gets rejected by the loop) of hundreds of Hz (Tyson 2022\cite{tyson2022}, chapter 2). This extensive experience in astronomy provides valuable insights for implementing \ac{RTAO} in high-energy laser systems.\\
        To leverage this expertise, we partnered with the \ac{SCExAO} team at the Subaru Observatory on Maunakea, Hawaii\cite{jovanovic2015}. The \ac{SCExAO} team, developers of the open source \ac{CACAO} software\cite{guyon2018, guyon2020, deo2024}, supported us in adapting their technology to our laser requirements on various levels.

    \subsection{Design Considerations}
        Implementing RTAO in high-energy laser systems presents several unique challenges that differ from those encountered in astronomy. These differences necessitate specific adaptations in our approach:\\
        
        \paragraph{Light Source}
        One major difference is the light source for sensing. In astronomy, continuous light from natural or artificial guide stars (Tyson 2022\cite{tyson2022}, chapter 3.3) is typically available. However, high-energy, high-intensity laser systems operate with pulsed beams, with intervals of no light that can last for minutes during shot preparation.\\
        One way to address this is to use a continuous pilot beam that co-propagates with the main beam, thus sampling the air turbulence and providing the \ac{RTAO} loop with a signal to act upon. Like that, the loop can compensate the dynamic aberrations in the beamline up to shot delivery. The main concern using this technique is the protection of the sensor from the high fluence that occurs on shot. To mitigate this, the pilot beam can either be blocked a few milliseconds prior to the shot using a fast shutter, or the beam can be passively separated - e.g. by employing an off-spectral pilot beam that features a wavelength that can be separated from the main pulse using a dichroic mirror and filters, which we decided to du at Apollon.\\
        
        \paragraph{Wavefront Sensor}
        The choice of \ac{WFS} also required adaptation. While \acp{PWFS} are commonly used in astronomy due to their ability to deal with photon noise when running on dim light sources, as well as high spatial resolution and computationally cheap evaluation (Tyson 2022\cite{tyson2022}, chapter 5.3), the \ac{PWFS} faces limitations in its ability to deal with non-common path aberrations. This is typically no big concern in astronomical instrumentation as the beam diameters remain small after splitting the light between the \ac{WFS} and the scientific camera. In our field, however, the sampling is done at the largest beam diameters after the beam is amplified in order to prevent damaging optics. The successive imaging and demagnification system usually generates significant amounts of non-common path aberrations, which renders \acp{PWFS} unfeasible for this application.\\
        For this reason, other \ac{WFS} schemes that cope well with this issue have to be used while carefully tweaking the evaluation pipeline to still feature sufficient speed and throughput for \ac{RTAO} applications. One example would be using a \ac{SHS}, which features a large dynamic range, and implement a suitably fast evaluation routine, as done in this project.\\
        
        \paragraph{Deformable Mirror}
        Another critical consideration was the selection of the \ac{DM}. In astronomical instrumentation, small, fast, high-resolution \ac{MEMS} mirrors are commonly used as they greatly fit this application \cite{jovanovic2015, lozi2024}.\\
        However, the \ac{LIDT} of these \acp{DM} is too low for high-energy, high-intensity lasers, as the thin membranes cannot withstand the stress of highly reflective dielectric coatings. At the same time, this technology cannot be scaled up easily to larger diameters at reduced fluence.\\
        To our knowledge, the only \ac{DM} technology that allows to accommodate both specialized coatings and larger diameters are piezoelectric ones. Standard piezoelectric bimorphs feature a good balance of size, speed, stroke and high damage threshold, while remaining relatively affordable\cite{bonora2016}.\\
        
        \paragraph{\ac{RTAO} Controller}
        The demands to the computation hardware are easily derived from the requirements for latency (see Sec. \ref{sec:system-modeling}) and the number of control modes and quickly exceeds the performance of a \ac{CPU}-based system. The two main competitors for high-speed \ac{RTAO} control hardware are \ac{FPGA}\cite{kudryashov2018, surendran2018, rukosuev2020} and \ac{GPU}\cite{guyon2018, truong2008} based systems. Both deliver high throughput and low latency computation to achieve the required performance.\\
        In the laser community, \ac{AO} controllers were usually non-real-time application that exclusively ran on the computers \ac{CPU}. Despite \acp{FPGA} potentially featuring lower latency, the increased and hardware-specific development efforts rule this option out as a community project. For this reason, using open-source \ac{GPU} code on a \ac{RTC} is a more suitable approach.

\section{Implementation} \label{sec:artao}
    In this section, we provide details of the implementation of the \ac{ARTAO} system. The setup is based on the design considerations that we discussed in the last section.
    
    \subsection{Pilot Beam Approach}
        For \ac{ARTAO}, we chose the passive off-spectral pilot beam scheme. Our previous investigations showed that a wavelength of 905\,nm is easy to separate from the main spectrum (730\,nm - 890\,nm) using dichroic mirrors and filters, while featuring a transmission of 10\% up to the 1\,PW compressor entrance, which is sufficient to saturate the \ac{WFS} at an injected beam power of 20\,mW. Moreover, the passive approach is easier to implement and more robust as there is no risk of damaging the \ac{WFS} due to shutter failure.\\
        We injected the single-mode-fiber-coupled 905\,nm laser diode after the second amplifier where the beam was still small and not impaired by air turbulence. The pilot beam was then separated right before the vacuum tank of the 1\,PW compressor (see section \ref{sec:integration} for details on the sensor setup).
        
    \subsection{Real-Time Controller} \label{sec:real-time-controller}
        As this is, to the best of our knowledge, the first time that a fully custom \ac{RTAO} loop was implemented in an ultra-intense laser system, reusability in other systems was a concern. For that reason, we decided to adapt the open source \acf{CACAO} code, developed by the \ac{SCExAO} team, due to its robust features, maturity, and its use of \acp{GPU} to achieve high-throughput, low latency operation. The architecture and features of \ac{CACAO} are described in detail in the original publications by the \ac{SCExAO} team\cite{guyon2018, guyon2020, deo2024}. In brief, \ac{CACAO} can be described as a framework where computations are split into microservices. These processes are triggered by image stream updates and thus form a fully customizable \ac{AO} pipeline.\\
        For \ac{ARTAO}, we extended \ac{CACAO} by implementing hardware interfaces, custom \ac{WFS} referencing and evaluation functions and a vectorized \ac{WF} management to avoid unnecessary reshaping operations.\\
        The hardware that we used to run \ac{CACAO} was inspired by the KalAO\cite{hagelberg2020} system that featured specifications that were comparable to what we intended to achieve. The \ac{RTC} was assembled solely of consumer grade components (see appendix \ref{sec:appendix-rtc} for a full list of components) and can be compared to a high-end gaming computer. As such, the assembly was relatively cheap compared to the high-speed camera and the \ac{DM} while easily delivering the required performance (see Sec. \ref{sec:system-modeling} for requirement estimation).

    \subsection{DM Selection}
        As discussed before, the requirements for speed and damage threshold significantly reduce the number of choices concerning \ac{DM} technology, where piezoelectric \acp{DM} are currently the most feasible options. While placing a large \ac{DM} after the amplifiers as close to the \ac{WFS} as possible would be the most unproblematic solution due to more direct \ac{DM} imaging and less pre-compensation (foreshadowing, see Sec. \ref{sec:future-work}), the increased size does not only mean increased costs, but also a lower resonance frequency for bimorph \acp{DM}. Piezostack \acp{DM} would avoid this issue, but feature a significantly lower dynamic range.\\
        For these reasons, we decided to place a smaller bimorph \ac{DM}, optimized for a 55\,mm beam diameter, prior to the second last amplifier and image it to the wavefront sensor instead. The \ac{DM} does therefore pre-compensate the dynamic aberrations in the main amplifier.\\
        In this work, we were using a bimorph with 60\,mm clear aperture and 96 actuators in an annular arrangement, optimized for Zernike stroke at an angle of incidence of 0°-10° (\textit{DM9660} by Dynamic Optics, Italy). The characterization is shown in section \ref{sec:testbench-performance}.

    \subsection{Wavefront Sensor}
        For the \ac{WFS}, we constructed a custom \ac{SHS} using a \ac{MLA} with a pitch of 250\,µm and a focal length of 11\,mm and a high-speed PCIe camera (\textit{CB013MG-LX-X8G3-TG}, xiB-64 series by XIMEA), capable of streaming frames directly to the \ac{RTC}s memory (or even to the \ac{GPU}, if it features GPUDirect RDMA) at up to 7\,kHz for a \ac{ROI} of 5$\times$5\,mm\textsuperscript{2} (384$\times$384\,px\textsuperscript{2}) at a negligible latency. We used the taped glass version of the camera in order to remove the cover glass from the chip as this would cause interference patterns when a coherent, monochromatic light source, such as our \ac{CW} pilot beam, is employed.\\
        We determined the distance of the \ac{MLA} to the camera chip by introducing know amounts of tilt and observing the motion of the centroids on the camera image. The exact knowledge of this distance enables quantitative wavefront measurements relative to a reference. The reference was kept relative by design as the beam optimization at Apollon is done using a standard procedure with a large, slow mechanical \ac{DM} (\textit{ILAO180}, Imagine Optic), which provides the target \ac{WF} to be stabilized.\\
        For the evaluation, we implemented two routines in \ac{CACAO}: a referencing routine, generating a fixed \ac{WF} pupil with reference positions and a \ac{GPU}-based routine that uses this reference to calculate the \ac{WF} from the camera image. The latter is logically a three-step process: first, the image is convolved with a Gaussian kernel, fitting the standard deviation of the spots of the \ac{SHS}, in order to increase robustness against noise. Second, a parabola is fitted to the brightest pixel and it's closest neighbors of each smoothed spot in order to retrieve the fractional shift of the spot\cite{poyneer2003}. And third, the retrieved slopes are multiplied with a matrix that translates the gradient vector to a \ac{WF} vector.\\
        Practically, we used an adaptive windowing approach, where only the relevant pixels around each spots are copied to the \ac{GPU}. since the standard deviation of the focal spots (13.9\,\textmu m) nicely matched the camera pixel size (13.7\,\textmu m) in our case, the algorithm uses a kernel size of 3$\times$3 pixel. Under the assumption that the spot position did move less than half a pixel since the last frame, 6$\times$6 pixels, excluding the corner pixels, need to be streamed to the GPU.\\
        The entire operation is completed within approximately 60\,µs, well within the latency budget.

\section{Characterization and Testing} \label{sec:characterization}
    \subsection{System Modeling} \label{sec:system-modeling}
        \begin{table*}[b]
            \centering
            \begin{tabular}{c|l|c|c|c}
                Parameter & Description & Target & Testbench & Apollon \\\hline
                $T_\mathrm{exp}$ & \ac{WFS} exposure time, reciprocal of max. framerate & 100\,us & 1/(7\,kHz) & 1/(7\,kHz) \\
                $T_\mathrm{trans}$ & frame transfer time, reciprocal of max. framerate & 100\,us & 1/(7\,kHz) & 1/(7\,kHz) \\
                $G$ & control feedback gain & 0.5 & variable & 0.27; 0.35\\
                $T_\mathrm{c}$ & computation + driver latency, determined experimentally* & 1\,ms & 1.2\,ms & 1.0\,ms \\
                $T$ & control period, reciprocal of control frame rate & 1/(1\,kHz) & 1/(3.6 kHz) & 1/(2.6 kHz)\\
                $\lambda$ & leakage multiplier per step & 0.999 & 0.999 & 0.99 \\
                $T_\mathrm{DM}$ & \ac{DM} rise time (10\%-90\%), determined experimentally & 500\,µs & 486\,µs & 486\,µs\\
            \end{tabular}
            \caption{The parameters used in the model of the \ac{ARTAO} loop for the estimation, on the testbench and in the Apollon beamline itself.\\
            *$T_\mathrm{c}$ is dominated by the latency of the \ac{DM} driver. While this is hard to separate from other effects, measurements indicate a delay of approx. 0.7\,ms (see Sec. \ref{sec:testbench-performance})}
            \label{tab:SISO-params}
        \end{table*}
        \begin{figure*}[h]
            \centering
            \includegraphics[width=\textwidth]{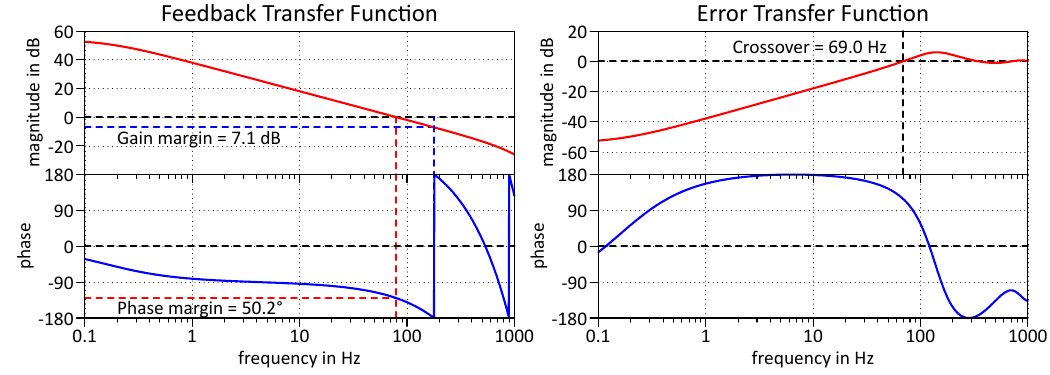}
            \\\vspace{-0.2cm}
            \caption{The Bode plot of the feedback transfer function $F(s)$ (left) and the error transfer function $R(s)$ (right) according to Eq. \eqref{eq:error-transfer-function}, using the parameter values from the "Target" column in table \ref{tab:SISO-params}.}
            \label{fig:target-bode-plot}
        \end{figure*}
        In order to verify the system's performance at a later point, we approximated the loop as a single-input-single-output system and constructed a transfer function in the Laplace domain (Tyson 2022\cite{tyson2022}, chapter 8.2). In the following, we briefly describe the components of the feedback transfer function, where the formulae are listed after the enumeration:
        \begin{enumerate}
            \item{\ac{WFS} Exposure:} $H_\mathrm{exp}(s)$\\
            Modeled as an integrator normalized to the exposure time $T_\mathrm{exp}$, minus another normalized integrator delayed by $T_\mathrm{exp}$, effectively simulating the average response within the exposure window.
            \item{Frame transfer:} $H_\mathrm{trans}(s)$\\
            We represented the delay due to the \ac{WFS} frame transmission Represented as a simple delay corresponding to the max. possible frame rate - approximately $T_{trans} = $135\,us in our case.
            \item{Controller Response:} $H_\mathrm{RTC}(s)$\\
            The control unit was emulated by a discrete sampling (chain of dirac impulses) every control cycle period $T$, coupled with a delay $T_\mathrm{c}$, resembling the computation time and driver latency, and a gain factor $G$.
            \item{Leakage:} $H_\mathrm{leak}$\\
            As our loop ran using a leaky integrator, i.e. a small portion of the integrated signal is "thrown away" in order to avoid the accumulation of errors, the leakage per control step needed to be included in the model. For this, we modified the transfer function of an exponential decay in a way that we could use the per-step leakage multiplier $\lambda$ to determine how much of the control signal is discarded every iteration.
            \item{Driver Hold:} $H_\mathrm{DRV}(s)$\\
            The \ac{DM} driver itself was modeled by a simple zero order hold for the control cycle period $T$. Considering the control signal being a chain of dirac pulses with variable amplitude, this is again an averaging over the control period $T$.
            \item{\ac{DM} Response:} $H_\mathrm{DM}(s)$\\
            The speed of the bimorph \ac{DM} is limited by the charge time of the electrodes. For this reason, we approximate the rise time by a linear ramp. Taking the step-like signal from the driver, this is yet another averaging transfer function. The rise time $T_\mathrm{DM}$ was extracted from measurements (see next section.)
        \end{enumerate}
        In formulae, these contributions equal:
        \begin{align*}
            H_{\mathrm{exp}}(s) &= \frac{1}{sT_\mathrm{exp}} - \frac{e^{-sT_\mathrm{exp}}}{sT_\mathrm{exp}} = \frac{1-e^{-sT_\mathrm{exp}}}{sT_\mathrm{exp}}\\
            H_\mathrm{trans}(s) &= e^{-sT_\mathrm{trans}}\\
            H_\mathrm{RTC}(s) &= G \cdot \frac{e^{-sT_\mathrm{c}}}{1-e^{-sT}}\\
            H_\mathrm{leak}(s) &= \frac{sT}{sT - \ln(\lambda)}\\
            H_\mathrm{DRV}(s) &= \frac{1-e^{-sT}}{sT}\\
            H_{\mathrm{DM}}(s) &= \frac{1-e^{-sT_\mathrm{DM}}}{sT_\mathrm{DM}}\\
        \end{align*}
        The parameters are listed in table \ref{tab:SISO-params}.\\
        The corresponding error transfer function of the loop is
        \begin{align}
            R(s) = \frac{1}{1+F(s)},\label{eq:error-transfer-function}
        \end{align}
        where $F(s)$ is the feedback transfer function, which in turn is the product of all transfer functions discussed above. The corresponding magnitude Bode plot (amplitude response) at frequency $\omega$ is then obtained by evaluating $R(i\omega)$.
        Prior to the design of \ac{ARTAO}, we evaluated $R(s)$ for a set of different parameters to aid the selection of the hardware. In table \ref{tab:SISO-params}, we listed the set of target parameters that yielded a crossover frequency of approx. 70\,Hz (cutoff frequency of the measured turbulence) with suitable gain and phase margins (see \cite{tyson2022}, chapter 8.2.9), while being realistically achievable. The corresponding Bode plots are shown in Fig. \ref{fig:target-bode-plot}.\\
        We will use the same model with different parameter sets in the next section to validate our understanding of the system with experimental data.

    \subsection{Testbench Setup}
        \begin{figure}[H]
            \centering
            \includegraphics[width=1\linewidth]{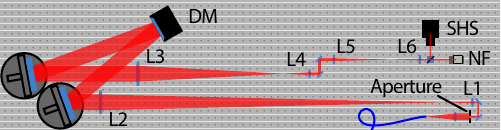}
            \\\vspace{-0.3cm}
            \caption{Schematic of the test-bench setup, to scale.}
            \label{fig:testbench-setup}
        \end{figure}
        To evaluate the functionality and performance of the \ac{ARTAO} system, we set up a testbench with a simplified configuration (see Fig. \ref{fig:testbench-setup}). It consisted of a single beampath where we collimated the pilot source after exiting the fiber, an aperture fixing the beam size, an imaging beam expander to the \ac{DM} and an imaging system to the \ac{SHS} afterwards. Although we did not implement a physical turbulence generator to emulate the operational conditions, we utilized \ac{CACAO}’s internal turbulence injection to introduce disturbances to the \ac{DM}.\\
        The primary objectives of this testbench setup were to verify the system's functionality, characterize the \ac{DM}, measure system latency, and determine the frequency response.
    
    \subsection{Performance Evaluation} \label{sec:testbench-performance}
        We conducted several performance tests to validate the system.

\begin{figure}[H]
            \centering
            \includegraphics[width=1\linewidth]{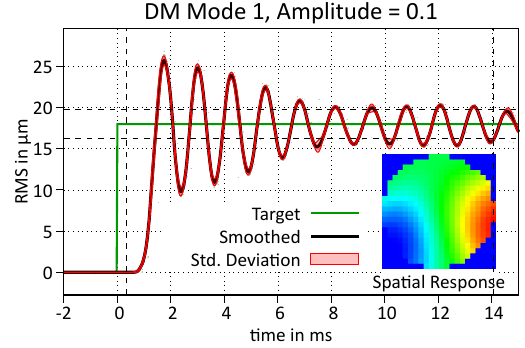}
            \includegraphics[width=1\linewidth]{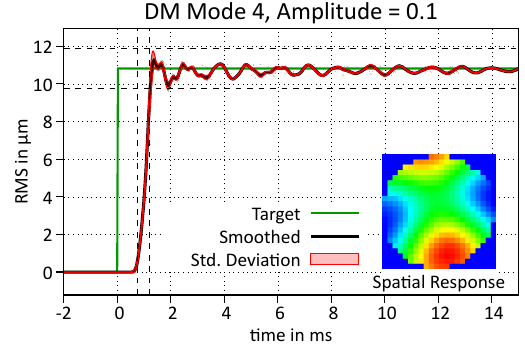}
            \caption{Step response of two \ac{DM} modes over time (open loop). The dashed lines indicate the 10\%-90\% levels and corresponding settling times. Top: first mode, featuring severe ringing due to the mechanical DM properties. Bottom: fourth mode with a regular settling behavior. All modes from this order upwards feature a comparable settling behavior (see Fig. \ref{fig:dm-all-responses}.}
            \label{fig:dm-response-singles}
        \end{figure}

\begin{figure}[H]
            \centering
            \includegraphics[width=1\linewidth]{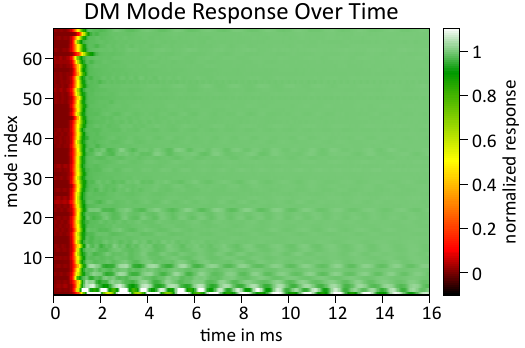}
            \caption{Step responses of a full set of mirror modes over time.}
            \label{fig:dm-all-responses}
        \end{figure}

        \paragraph{\ac{DM} Latency and Rise Time:}
        Using a strobing technique, we evaluated the \ac{DM} latency and rise time. By repeatedly poking a selected \ac{DM} mode a couple of hundred times while varying the delay with respect to the camera frame, we generated a corresponding response curve (see Fig. \ref{fig:dm-response-singles} for single-mode example curves), which is the \ac{DM} response, convolved with the camera exposure timeframe, and delayed by the driver delay, the camera delay and the \ac{WFS} evaluation time. From this, we extracted both the latency between the command and the first available wavefront of the onset of the risetime and the 10\%-90\% rise time of the \ac{DM} itself. We then repeated process for all control modes of the control matrix (see Fig. \ref{fig:dm-all-responses}). The rise time is fairly consistent across all modes except for the lowest three ones, roughly corresponding to Tip/Tilt and Defocus. When excluding these three, the rise time is at 486$\pm$50\,µs, following a delay of 780$\pm$49\,µs (including driver delay, camera exposure, frame transfer and \ac{WFS} evaluation).\\
        The lowest three modes showed notable ringing due to the \ac{DM} mechanics. This ringing significantly increased the settling time to 13\,ms, 8.3\,ms and 1.4\,ms, respectively. While this is part of a separate project, this clearly indicates the advantage of a separate pointing control system if no \ac{DM} with better temporal characteristics on these modes is available.

\begin{figure}[H]
            \centering
            \includegraphics[width=1\linewidth]{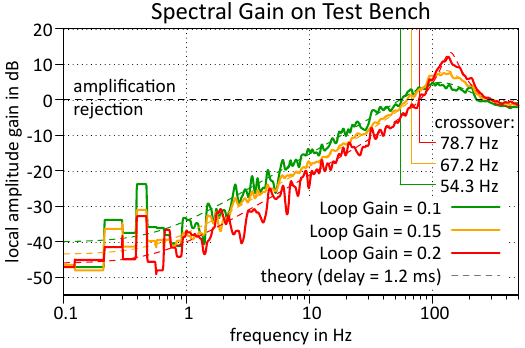}
            \caption{The AO loop gain over the frequency for different feedback gains. The dashed curves are the gains of the corresponding model, i.e. the magnitude of the error transfer function.}
            \label{fig:testbench-gain}
        \end{figure}

        \paragraph{System Frequency Response:}
        We measured the system's frequency response using \ac{CACAO}’s internal turbulence emulator, which introduced disturbances on the \ac{DM} by sweeping a Kolmogorov screen \cite{kolmogorov1968} across the actuator matrix at a constant velocity. By measuring the local wavefront jitter in both open and closed loop configurations, applying spectral smoothing, and determining the ratio of the two for different stable loop gains, we experimentally measured the frequency response of the system (see Fig. \ref{fig:testbench-gain}).\\
        At this point, we used the model for the error transfer function (Eq. \eqref{eq:error-transfer-function}) as a fit model, where the loop delay was the fit parameter. At a delay of 1.2\,ms, we found excellent agreement between the model and the experimental data.\\
        In the lowest frequency range, the deviation of the data from the simulated curves increases, while also featuring some correlation between the measurements. The latter can be attributed to the periodic nature of the Kolmogorov screen that we used to disturb the loop, while the magnitude of the errors are caused by the limited recording duration (approx. 13.65\,s in 12 image cubes, 4096 frames each), which leads to poor statistics in the low frequency domain.

\section{Integration with Apollon Laser System} \label{sec:integration}
    \subsection{Implementation Details}
        \begin{figure*}[h]
            \centering
            \includegraphics[width=1\linewidth]{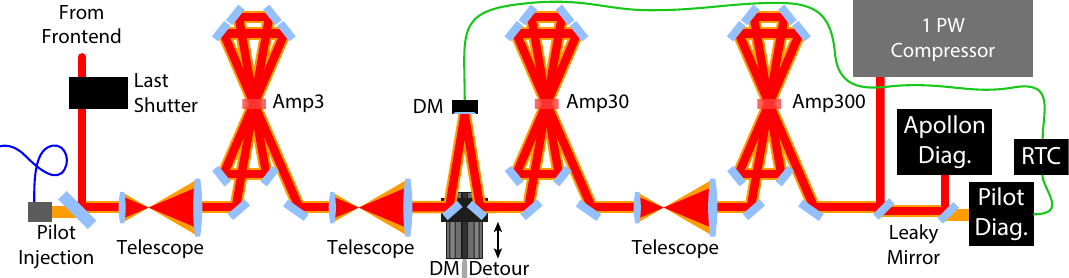}
            \caption{Schematic setup of \ac{ARTAO} in the Apollon laser chain}
            \label{fig:artao-setup}
        \end{figure*}
        We performed the integration of the \ac{ARTAO} system into the Apollon laser system in a way that did not require significant changes in the main beamline that could potentially interfere with regular operation. In the following, we will briefly describe the changes we made. The setup is shown schematically in Fig. \ref{fig:artao-setup}.\\
        We injected the 905\,nm \ac{CW} pilot beam right after the last Pockels cell shutter of Apollon, located between the second and third amplifiers. Using a reflective fiber collimator, the beam featured a diameter slightly above the diameter of the main beam. We aligned the polarization using a wave plate and injected the pilot through a dichroic mirror. The regular beam alignment references were used to ensure co-propagation of both the pilot and main beams.\\
        In order to enable easy bypassing of the \ac{DM}, we integrated a beam detour prior to the fourth amplifier, where the beam diameter is 55\,mm, ensuring a low fluence to prevent any potential damage to the \ac{DM}. The detour injection- and pickup mirrors were placed on a linear stage and could be inserted and removed whenever necessary.\\
        \begin{figure}[H]
            \centering
            \includegraphics[width=1\linewidth]{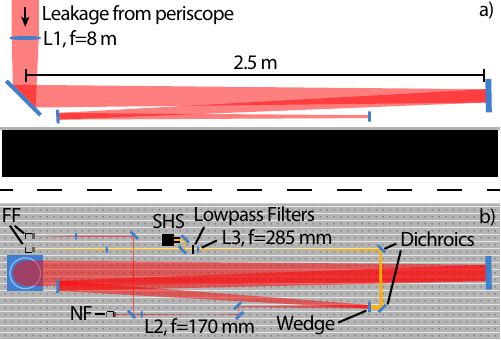}
            \caption{Sketch of the diagnositc setup prior to the 1\,PW compressor in side- (a) and top view (b). The main beam path is shown in red, while the pilot beam path is indicated in orange.}
            \label{fig:sensor-setup}
        \end{figure}
        For the sensor setup, we implemented a dedicated beam sensor configuration right before the entrance of the 1\,PW compressor (see Fig. \ref{fig:sensor-setup}). Leakage light from the main beam was focused by an f = 8\,m lens and then collimated. A dichroic mirror, followed by spectral filters, separated the pilot and main beams, allowing for independent diagnostics of each. The fast \ac{SHS} was placed in the pilot beam arm in the image plane of the \ac{DM} and was complemented by a \ac{FF} camera. Regular beam diagnostics, including \ac{NF}, \ac{FF}, and a slow \ac{SHS}, were placed in the main beam arm, where the image plane of the \ac{WFS} matched that of the \ac{DM} as well to enable direct comparison between the \acp{WF} of the two beams.\\
        The \ac{RTC} for ARTAO was connected to the fast \ac{WFS} via a 5\,m PCIe copper cable and positioned close to the setup. We used a 50\,m long CAT6 ethernet cable to directly connect the \ac{RTC} and the \ac{DM}, both of which are located in different rooms within the infrastructure.

    \subsection{Pilot Beam Correlation}
\begin{figure}[H]
            \centering
            \includegraphics[width=1\linewidth]{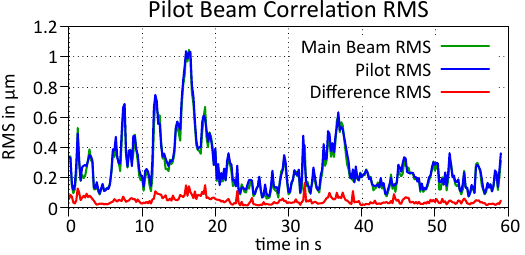}
            \caption{\ac{RMS} of the main- and the pilot beam \ac{WF}, as well as the difference between the two, over a time frame of one minute. The beam was actively disturbed using a hot air source for this measurement.}
            \label{fig:pilot-correlation-rms}
        \end{figure}

        Using a pilot beam introduces uncertainties due to the differences in frequency and potentially alignment. In order to ensure that running an \ac{RTAO} loop on the pilot beam benefits the real beam, we investigated the correlation between the two beams. To do so, we simultaneously recorded \acp{WF} on both beams using the diagnostic setup.\\
        \begin{figure}[h]
            \centering
            \includegraphics[trim=0 0 0 20, clip, width=1\linewidth]{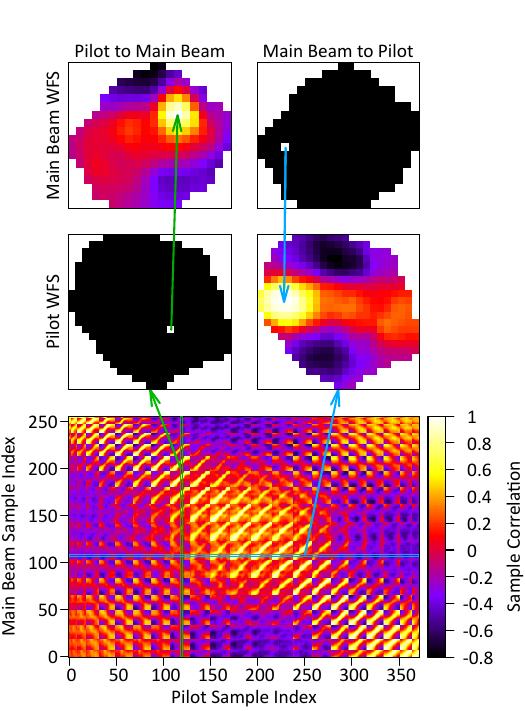}
            \caption{Bottom: the sample-wise correlation matrix between the \ac{WFS} of the main beam and the pilot beam over the recorded sequence without tilt and mean \ac{WF}. Top: example correlation of the main beam \ac{WF} to a randomly picked location of the pilot \ac{WF} (left) and vice versa (right). Point pairs for a transformation fit can be extracted from the locations of max. correlation.}
            \label{fig:correlation-mapping}
        \end{figure}
        To determine the mapping from one \ac{WFS} to the other, we constructed a sample-wise correlation matrix from the recorded sequences, where we subtracted Tip/Tilt and the mean \ac{WF} beforehand (see Fig. \ref{fig:correlation-mapping}). This matrix illustrates how well each \ac{WF} sample in one \ac{WFS} space correlates to a specific pixel in the other \ac{WFS} space. We extracted the positions of maximum correlation, generating a set of point pairs, which we used to fit an affine transformation. This transformation allowed us to convert \acp{WF} between the two spaces using linear interpolation. An example frame is shown in Fig. \ref{fig:correlation-frame-example}, with an animated version in the supplementary material of this publication.\\
        \begin{figure}[h]
            \centering
            \includegraphics[width=1\linewidth]{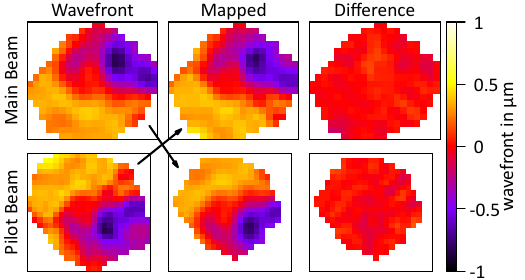}
            \caption{Example of a mapped \ac{WF} between the main beam \ac{WFS} (top row) and the pilot beam \ac{WFS} (bottom row), where the first column is the raw \ac{WF}, the second one is the mapped \ac{WF} from the other \ac{WFS}, respectively, and the last column is the difference between the two. Note that the main beam is smaller than the pilot beam, which is why its mapped \ac{WF} is smaller in the pilots \ac{WFS} space.}
            \label{fig:correlation-frame-example}
        \end{figure}
        The analysis revealed excellent correlation between the \acp{WF} of the pilot and main beams, with errors being both of high spatial order (see Fig. \ref{fig:correlation-frame-example}, right column) and of significantly smaller amplitude than the full beam disturbance (less than 20\%, see Fig. \ref{fig:pilot-correlation-rms}). Potential sources of these errors include imperfect synchronization (the fast \ac{SHS} was not triggered, estimated at approx. 20\,ms accuracy in frame selection), errors in the fitted transformation, and linear interpolation inaccuracies. While the precise attribution of these errors to the sources remains uncertain, we consider the overall error magnitude to be sufficiently small to validate the pilot beam concept for this application - especially since the Strehl ratio deficit scales quadratically with the \ac{RMS} for small aberrations.

    \subsection{Short-term Stabilization}
        \begin{figure}[h]
            \centering
            \includegraphics[width=\linewidth]{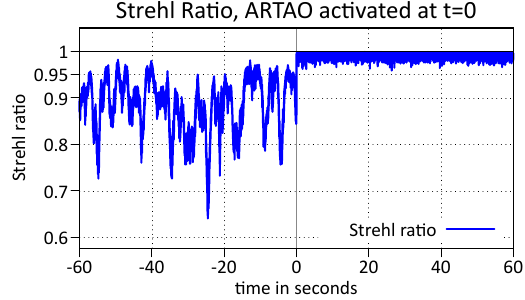}
            \includegraphics[width=\linewidth]{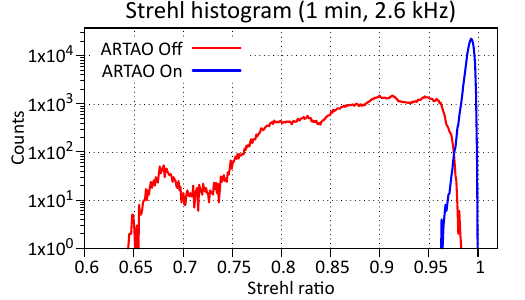}
            \vspace{-0.35cm}
            \caption{Top: time series measurement of the \correction{calculated }{}Strehl ratio (compared to the reference \ac{WF}\correction{}{, calculated via the \ac{FFT} of the measured \ac{NF}}) of the pilot beam, where \ac{ARTAO} is activated at t=0. Bottom: the histogram of the corresponding data series.}
            \label{fig:artao-strehl}
        \end{figure}
        \begin{figure}[h]
            \centering
            \includegraphics[width=1\linewidth]{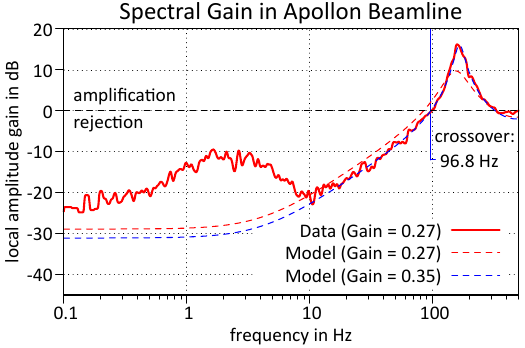}
            \caption{Recorded gain curve of ARTAO on the pilot beam \ac{WF} in the Apollon beamline under regular opeartion conditions, compared with two theoretical curves (dashed lines) with parameters from table \ref{tab:SISO-params}. The dashed red curve uses the parameters of the real-world loop, while we tweaked the feedback gain for the blue curve in order to match the data.}
            \label{fig:pilot-gain}
        \end{figure}
        To evaluate the short-term stabilization performance of the RTAO system, we conducted a series of experiments under regular operational conditions, (i.e. all beam housings in place and air conditioning systems active). The primary goal was to achieve optimal \ac{WF} stability by fine-tuning the loop parameters to the environment.\\
        In these experiments, we mainly adjusted the framerate of the \ac{WFS}, the Tip/Tilt gain and the modal gain curves. For each setting, we incrementally increased the feedback gain to the edge of stability. We then recorded a time sequence of \acp{WF} of one minute prior to closing the loop and continued to record for another minute. From the \acp{WF}, we calculated the Strehl ratio\correction{}{ via a \ac{FFT} of the measured \ac{NF}}, which we used as relative measure for the beam stability. This Strehl ratio obviously only applies if all static aberrations are removed up to the target area\correction{}{ and is thus an unrealistic value from the experiment point of view, but it remains useful to build some intuition on the performance of the \ac{AO} loop in terms of achievable stability}. The result is shown in Fig. \ref{fig:artao-strehl}. The loop exhibits excellent performance on this short time scale, increasing the minimum Strehl ratio from 0.65 to 0.96.\\
        Additionally, we calculated the spectral gain curve in order to compare the data to our model. The parameters that we used to obtain the theoretical curves are shown in the right column in table \ref{tab:SISO-params}. Two deviations meet the eye: first, the rejection is significantly worse than the model for frequencies lower than 10\,Hz. The reason for this is that we characterized the gain based on the actual recorded \acp{WF}, where it's portion that is not covered by the \ac{DM} modes remains uncorrected. Apparently, the spatial frequencies feature a maximum in the low frequency regime, where they are additionally more visible due to the otherwise large rejection of the loop. Secondly, plugging in the actual loop parameters into the model (dashed red curve in Fig. \ref{fig:pilot-gain}) does not yield a good overlap in the higher frequency regime anymore. This is likely due to the fact that we applied a modal gain curve to the real-world loop, featuring a significantly higher gain (factor 1.4) for the Tip/Tilt modes of the \ac{DM}, which is obviously not reflected by the model. With an empirical upwards correction of the loop gain (0.27 to 0.35 $\approx$ a factor of 1.3, which is in between the base gain and the Tip/Tilt gain), the model resembles the data well again (dashed blue curve in Fig. \ref{fig:pilot-gain}). The crossover frequency at the edge of stability was 96.8\,Hz.\\
        \begin{figure}[H]
            \centering
            \includegraphics[width=1\linewidth]{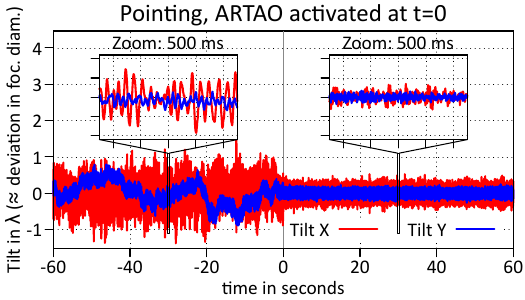}
            \vspace{-0.6cm}
            \caption{Time series measurement of the beam pointing of the pilot beam, where \ac{ARTAO} is activated at t=0.\correction{}{ The tilt X (red) and tilt Y (blue) curves represent the tilt-portion of the recorded \ac{WF}, relative to the reference, and is given in peak-to-valley in terms of the central wavelength of Apollon, roughly corresponding to the movement of the focal spot in focal spot diameters.} The inserts show a zoomed in portion to illustrate the fast oscillations in the beam pointing.}
            \label{fig:artao-pointing}
        \end{figure}
        Furthermore, we analyzed the data with respect to the pointing stability of the beam with and without \ac{ARTAO} (see Fig. \ref{fig:artao-pointing}). This plot demonstrates that the loop is able to achieve some level of compensation of the pointing instabilities, the limitations of the \ac{DM} and the loop bandwidth in general become relevant here. The pointing is dominated by an oscillation at around 35\,Hz, as shown in the zoomed insert in Fig. \ref{fig:artao-pointing}. This is a significant challenge considering the ringing of the first modes of the DM (see Fig. \ref{fig:dm-response-singles} and \ref{fig:dm-all-responses}) and the oscillations cannot be suppressed effectively, only yielding a Peak-to-Valley rejection of 50\% compared to the open loop. This underlines the necessity to implement a fast steering mirror to outsource this mode to a faster, more rigid system. While this is a project currently under its way at Apollon, this goes beyond the scope of this work.

    \subsection{Impact of Pumping the Amplifier Crystals}
        \begin{figure}[H]
            \centering
            \includegraphics[width=1\linewidth]{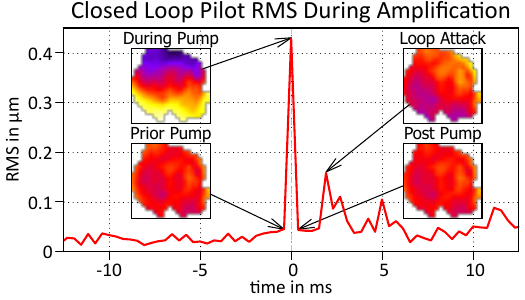}
            \caption{Measured \ac{RMS} of the closed loop pilot \ac{WF} during a pump event on the amplifiers. The inserts are the \acp{WF} at the times indicated by the arrows.}
            \label{fig:closed-loop-while-pumping}
        \end{figure}
        Although we refrained from running the loop while delivering amplified shots due to the risks involved (as discussed in section \ref{sec:future-work}), we investigated the impact of amplifier pumping on the loop stability: we flashed the amplifiers without a frontend beam while the \ac{ARTAO} loop was closed on the pilot beam in order to observe any potential effects on the system's performance.\\
        The results, presented in Fig. \ref{fig:closed-loop-while-pumping}, show the \ac{RMS} of the measured \ac{WF} over time, while the \ac{ARTAO} loop was closed. In the center of the plot, the amplification event occurs. At this instant, the \ac{SHS}'s spots became saturated due to the amplifier gain at 905\,nm. While this saturation did not pose a risk to the camera, it resulted in a presumable erroneous detection of \ac{WF} tilt. Approximately 2\,ms later, the loop reacted to the tilt measurement, attempting to compensate, which led to a noticeable but temporary reaction in the \ac{WF}. However, the disturbance was insufficient to destabilize the loop, and the \ac{WF} quickly returned to its optimal state within a few tens of milliseconds - far below the repetition rate of Apollon or even the frontend of the facility.\\
        These observations indicate that while amplifier pumping does momentarily affect the WF measurement, the RTAO system is robust enough to quickly recover, maintaining overall stability in the short term.

\section{Challenges and Future Work} \label{sec:future-work}
    While the results described in the previous sections demonstrates significant progress, several challenges remain before \ac{RTAO} can be considered feasible for high-power laser systems. This section outlines the key issues encountered during testing and proposes steps to address them.

    \subsection{Long-Term Stability Issues}
        \begin{figure}[H]
            \centering
            \includegraphics[width=1\linewidth]{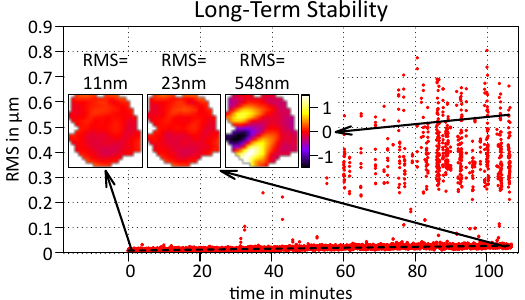}
            \\\vspace{-0.3cm}
            \caption{The \ac{WF} \ac{RMS} of the pilot beam under closed loop operation over an extended time frame. The insert plots are three selected \ac{WF} frames from stable conditions in the beginning (left) and the end of the recording (center), as well as from a period of instability (right).}
            \label{fig:lt-stability}
        \end{figure}
        A critical limitation of the \ac{ARTAO} system arises when considering long-term stability. Fig. \ref{fig:lt-stability} shows the evolution of the \ac{WF} \ac{RMS} of the pilot beam (relative to the initial reference) over nearly two hours of closed-loop operation. While the loop remains stable for the first 30 minutes, isolated instability events occur after this point, increasing in frequency until continuous instability streaks are observed beyond the one-hour mark. The baseline stability of the wavefront also degrades slightly over time.\\
        These instability events pose a significant risk for operation, as large \ac{DM} strokes during these phases can create hotspots in the beam, potentially damaging system optics during shot delivery. The unstable wavefronts exhibit wave-like structures (Fig. \ref{fig:lt-stability}, right inset) oscillating \correction{}{rapidly} over time, strongly suggesting lateral misregistration between the \ac{WFS} and \ac{DM}\cite{berdeu2024}. This is likely due to environmental changes\correction{}{, e.g. temperature or humidity} deforming the beamline slightly and thus shifting the \ac{DM} image on the \ac{WFS} over time.\\        
        This hypothesis was confirmed when minor\correction{}{ manual} adjustments to a \correction{}{randomly picked }steering mirror outside the \ac{DM} conjugation plane \correction{}{between the \ac{DM} and the \ac{WFS} realigned the \ac{DM} image position, which in turn} restored stability. Thus, implementing a tracking routine to maintain lateral beam alignment on the \ac{WFS} is a promising countermeasure. This approach will require a sensitive analysis routine to detect beam drift early \cite{berdeu2024}.\\
        \correction{}{It should be noted, however, that other sources of instability are still worth consideration. This includes, e.g., a drop in illumination on the \ac{WFS}, software crashes and exceedingly fast vibrations. This implies that a robust machine safety framework is necessary in any case.}

    \subsection{Nearfield Intensity Variations}
        \begin{figure}[H]
            \centering
            \includegraphics[width=1\linewidth]{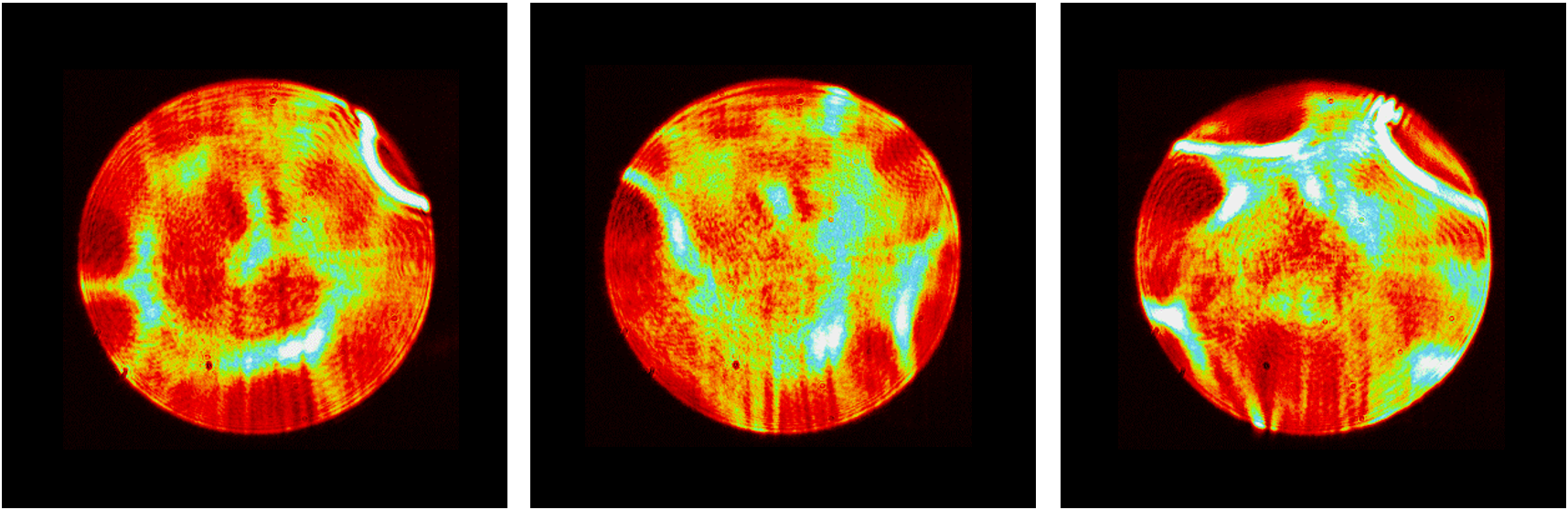}
            \caption{Imprints of artificially large \ac{DM} strokes onto the \ac{NF} fluence in a non-conjugate image plane\correction{}{, where each \ac{NF} corresponds to a different set of random actuator positions}.}
            \label{fig:np-imprints}
        \end{figure}
        Another issue arises from the coupling between the pre-compensated \ac{WF} and the \ac{NF} intensity. Aberrations introduced in planes not conjugated to the \ac{DM} lead to pre-compensation effects that couple the \ac{WF} to NF intensity variations. Depending on the magnitude of the aberrations, this coupling can generate hotspots, reducing the system's maximum supportable pulse energy. Artificial examples demonstrating this effect are shown in Fig. \ref{fig:np-imprints}, where deliberate large \ac{DM} strokes, approx. one order of magnitude higher than in regular operation, illustrate the potential for hotspot generation. Such amplitudes do occur for unstable loops\correction{}{, but could also occur for large amounts of pre-compensation, even if the loop remains stable. In this case, experiments may still be feasible at reduced energy, while the impact of the distorted \ac{NF} and thus the three-dimensional intensity distribution around the focal plane on the target interaction has to be considered for each experiment individually}.\\
        \correction{This}{The \ac{WF} to \ac{NF}} coupling is more pronounced for smaller beam diameters due to stronger \ac{WF} gradients for the same phase amplitude. Mitigation strategies therefore should focus on reducing pre-compensation stroke magnitudes. Assuming passive aberration mitigation strategies are optimized, two potential solutions emerge:
        \begin{enumerate}
            \item Fast Post-Compensating \ac{DM}: Implementing a fast, large diameter \ac{DM} after the amplifier for direct aberration post-correction. This requires significant hardware development as larger \acp{DM} are slower compared to smaller ones.
            \item Cascaded Compensation: Employing a fast pre-compensating \ac{DM} combined with a slower post-compensating \ac{DM} in the same conjugate plane to form a cascaded control system. This would share the correction load and reduce the required pre-compensation stroke.
        \end{enumerate}
        Both approaches require further investigation in future work\correction{}{, but in general, reducing and possibly eliminating the pre-compensation amplitude as far as possible is the favorable path}.

    \subsection{Implementation of Safety Mechanisms}
        As discussed previously, a malfunctioning \ac{AO} loop poses a severe hazard for high-power laser facilities, with potential damage to compressor gratings reaching millions of dollars, alongside significant system downtime. To safely implement \ac{RTAO} for shot delivery, robust safety mechanisms are essential. A recommended starting point involves developing a comprehensive interlock system:
        \begin{itemize}
            \item Loop-Internal Interlock: A monitoring subroutine that evaluates control loop parameters in real time. Safety thresholds, such as excessive actuator values or sudden changes in \ac{DM} shape, trigger an abort signal to the control system.
            \item Loop-External Interlock: An imaging system monitoring NF homogeneity in a critical optical plane. Violations of safety thresholds, such as localized intensity spikes or clipping, also trigger an abort signal.
            \item Pre-Shot \ac{DM} Freeze: Before each shot, the \ac{DM} is locked in its current position, halting the control loop. This interval must balance minimal dynamic aberration evolution and sufficient time for external interlocks to perform a final quality check.
        \end{itemize}

    \subsection{Integration into Laser Control System}
        In addition to ensuring machine safety, integrating the \ac{RTAO} system into the broader laser control system is essential for operability. While \ac{CACAO} provides a powerful and flexible platform for \ac{RTAO}, it requires in-depth knowledge of \ac{AO} concepts, software structure and available commands, making it unsuitable for daily laser operations, where operators need to manage many subsystems simultaneously.\\
        Developing an interface that abstracts common workflows in \ac{CACAO} into few intuitive commands is critical. This interface would allow operators to execute essential tasks without detailed knowledge of the underlying system. The integration effort is an essential step toward making \ac{RTAO} a practical tool for routine high-power laser operation.

\section{Conclusion} \label{sec:conclusion}
    This work represents the first implementation of an \ac{RTAO} system in a high-power laser, demonstrating its potential to significantly enhance beam stability. This advancement enables more repeatable laser shots, reduces experimental error bars, and supports higher repetition rates.\\
    The \ac{ARTAO} system relies on an off-spectral pilot beam. The match between the \acp{WF} of the beams was found to be excellent, ensuring continuous \ac{WF} control for pulsed laser systems. \ac{ARTAO} operates at an acquisition rate of 2.6\,kHz and a control speed of 1.3\,kHz, achieving a crossover frequency of 96.8\,Hz for local wavefront fluctuations. Short-term stabilization results indicate a guaranteed Strehl ratio of >0.96 for a statically fully corrected beam, compared to 0.62 without \ac{AO}, and robust performance during amplifier pumping events.\\
    The feasibility of \ac{RTAO} has therefore been established. However, several issues remain which need to be resolved before \ac{RTAO} can be used in daily operation: First, long-term stability is not yet given as beam drifts may lead to lateral misregistration over time, rendering the loop unstable. Second, robust machine safety mechanisms, including conjugation tracking systems and interlocks, are essential to prevent shot delivery with suboptimal beam quality, which could damage optics in the beam path. Third, simplified software interfaces are necessary as laser operators need to supervise many subsystems at once, requiring a certain level of abstraction. These next steps will be crucial for advancing \ac{RTAO} technology in large laser facilities.\\
    Once these issues are solved, \ac{RTAO} will addresses critical challenges in high-power laser operations, particularly stabilizing beam quality for experiments sensitive to laser parameters, such as laser wakefield acceleration or the usage of beams with orbital angular momentum. Large laser facilities, including glass-based high-energy systems, can particularly benefit by actively mitigating temperature-induced fluctuations, enabling shorter cooldown times and improved operational efficiency.\\
    In the long term, we expect \ac{RTAO} to contribute to advancements in the entire field, including applications such as laser particle accelerators, inertial fusion energy, and integrated solutions for commercial laser systems. This may revolutionize beam stability, becoming a new standard in high-power and high-energy laser facilities.

\section*{Acknowledgements}
    The authors would like to thank Vincent Deo, Kyohoon Ahn, Bartomeu Pou and Olivier Guyon from the \ac{SCExAO} team for their support with installing, operating and co-developing CACAO, which made this project possible in the first place.\\
    This project has received funding by the European Union’s HORIZON-INFRA-2022-TECH-01 call under grant agreement number 101095207.

\begin{acronym}
	\acro{AO}{Adaptive Optics}
    \acro{ARTAO}{Apollon Real Time Adaptive Optics}
    \acro{CACAO}{Compute And Control for Adaptive Optics}
	\acro{CW}{Continuous Wave}
	\acro{CPA}{Chirped Pulse Amplification}
    \acro{CPU}{Central Processing Unit}
	\acro{DM}{Deformable Mirror}
	\acro{FF}{Far Field}
    \acro{FFT}{Fast Fourier Transform}
    \acro{FPGA}{Field-Programmable Gate Array}
    \acro{GPU}{Graphics Processing Unit}
	\acro{GSI}{GSI Helmholtzzentrum für Schwerionenforschung}
    \acro{IFE}{Inertial Fusion Energy}
    \acro{LIDT}{Laser-Induced Damage Threshold}
    \acro{MEMS}{Micro-ElectroMechanical System}
	\acro{MLA}{Micro Lens Array}
    \acro{MVM}{Matrix-Vector Multiplication}
	\acro{NF}{near field}
	\acro{OAM}{Orbital Angular Momentum}
	\acro{OAP}{Off-Axis Parabolic mirror}
    \acro{OPCPA}{Optical Parametric Chirped Pulse Amplification}
	\acro{PHELIX}{Petawatt High-Energy Laser for heavy Ion eXperiments}
    \acro{PWFS}{Pyramid Wavefront Sensor}
	\acro{QWLSI}{Quadri-Wave Lateral Shearing Interferometer}
	\acro{RMS}{Root Mean Square}
    \acro{ROI}{Region Of Interest}
    \acro{RTAO}{Real-Time Adaptive Optics}
    \acro{RTC}{Real-Time Computer}
    \acro{SCExAO}{Subaru Coronagraphic Extreme Adaptive Optics}
	\acro{SHS}{Shack-Hartmann Sensor}
	\acro{STC}{Spatio Temporal Coupling}
    \acro{THRILL}{Technology for High-Repetition-rate Intense Laser Laboratories}
	\acro{WF}{Wavefront}
    \acro{WFS}{Wavefront Sensor}
\end{acronym}

\bibliographystyle{unsrtnat} 
\bibliography{reference}

\appendix
\section{Real-Time Computer Assembly} \label{sec:appendix-rtc}
    The components that we used to construct the \ac{RTC} that we used in this work are listed in Table \ref{tab:rtc-parts}. These are consumer grade components and are available at a relatively low price. The total cost for the \ac{RTC}, as for the time of writing, is approximately 3,000\,€ (compared to tens of thousands for the \ac{SHS} and \ac{DM}). The entire setup was heavily inspired by the \ac{RTC} of the KalAO system\cite{hagelberg2020}.\\
    It is noteworthy that all available PCIe extension slots are occupied in this setup, so further extentions would require a different choice of hardware:
    \begin{enumerate}
        \item \textbf{PCI\_E1:} PCIe 5.0x8 (From CPU), occupied by \ac{GPU}
        \item \textbf{PCI\_E2:} PCIe 5.0x8 (From CPU), occupied by PCIe camera card
        \item \textbf{PCI\_E3:} PCIe 4.0x4 (From Chipset), occupied by Ethernet card for \ac{DM}
    \end{enumerate}
    The RTC is running unter Ubuntu 20.04 LTS.
    
    \begin{table*}
        \centering
        \begin{tabular}{|r|l|l|}
            \hline
            \textbf{Component} & \textbf{Description} & \textbf{Manufacturer, Model} \\\hline
            Processor &	AM5 socket, 16 Cores, <5.5\,GHz & AMD Ryzen 9 7950X \\\hline
            Motherboard & AM5 socket, 2 PCIe 5.0 slots from CPU & MSI MPG X670E Carbon WIFI \\\hline
            GPU & 12\,GB GDDR6X, 10240 CUDA cores, 1725\,MHz & Palit GeForce RTX 3080Ti GameRock \\\hline
            RAM & 2x16\,GB DDR5-4800 & Kingston FURY Beast RGB 16\,TB (x2) \\\hline
            NVMe SSD & PCIe 4.0, 2\,TB & Kingston KC3000 \\\hline
            Network card & Intel 82576 chip, GBit Ethernet, Dual RJ45 Ports & \textit{Unknown} \\\hline
            Chassis & 19" rack mountable & SilverStone RM42-502-B \\\hline
            CPU cooler & Water cooling, 2x 240\,mm radiator & SilverStone SST-PF240-ARGB-V2 \\\hline
            Chassis fans & 2x 80\,mm high-throughput fans & Alphacool ES 80\,mm 800-6000\,rpm \\\hline
            Power unit & 1\,kW, gold standard & be quiet! Pure Power 11 FM 1000\,W \\\hline
            Hard drive & 8\,TB for mass storage & Seagate BarraCuda 8\,TB\\\hline
        \end{tabular}
        \caption{Component list for the \ac{RTC} used in this work.}
        \label{tab:rtc-parts}
    \end{table*}
\end{document}